\documentclass{elsart}

\usepackage{graphicx}

\usepackage{amssymb}

\begin{document}

\begin{frontmatter}



\title{Microscopic black hole detection in UHECR: the double bang signature}


\author[coim]{V. Cardoso\thanksref{grant1}}
\author[lip]{M.C. Esp\'{\i}rito Santo\thanksref{grant2}}
\author[lip]{M. Paulos}
\author[lip,ist]{M. Pimenta\corauthref{cor}}
\author[lip]{B. Tom\'e\thanksref{grant3}}
\address[coim]{CFC, Universidade de Coimbra,
P-3004-516 Coimbra, Portugal}
\address[lip]{LIP, Av. Elias Garcia, 14--1, 1000-149 Lisboa Portugal}
\address[ist]{IST, Av. Rovisco Pais, 1049-001 Lisboa, Portugal}
\corauth[cor]{pimenta@lip.pt, LIP, Av. Elias Garcia, 14--1, 1000-149 Lisboa Portugal.}
\thanks[grant1]{FCT grant SFRH/BPD/14483/2003.}
\thanks[grant2]{FCT grant SFRH/BPD/5577/2001.}
\thanks[grant3]{FCT grant SFRH/BPD/11547/2002.}

\begin{abstract}
According to recent conjectures on the existence of large extra dimensions
in our universe, black holes could be produced during the interaction of Ultra High
Energy Cosmic Rays with the atmosphere. However, and so far, the proposed 
signatures are based on statistical effects, not allowing 
identification on an event by event basis, and may lead to large uncertainties.
In this note, events with a double bang topology, where the  
production and instantaneous decay of a microscopic black hole (first bang) is followed,
at a measurable distance, by the decay of an energetic tau lepton (second bang) are
proposed as an almost background free signature. 
The characteristics of these events and the capability of large cosmic ray experiments to 
detect them are discussed.
\end{abstract}

\begin{keyword}
\sep extra-dimensions \sep black holes \sep UHECR \sep EAS \sep 
neutrinos \sep taus \sep double bang \sep Auger \sep EUSO \sep OWL

\PACS 04.50.+h \sep 04.70.-s \sep 13.15.+g \sep 96.40.Pq 
\end{keyword}
\end{frontmatter}

\section{Introduction}
\label{sec:introd}

Recent attempts to solve the hierarchy problem rely on the existence of extra dimensions 
in our universe, thereby lowering the fundamental Planck scale down to TeV energies \cite{Hamed}. 
In such scenarios, gravity should get stronger at shorter distances, and therefore 
new phenomena should arise in TeV-scale experiments.
One of the most interesting of these new phenomena is the production of black holes
in collisions where the centre-of-mass energy is higher than 1 TeV \cite{Bhprod}.
In this context, Ultra High Energy Cosmic Rays (UHECR) stand out naturally as the best candidates
for black hole production, through the interaction with matter in the atmosphere \cite{Extradim}. 

The detection of events signaling black hole production, as well as their non-observation,
would be of extreme importance, allowing to put constraints on the number of extra dimensions and on the 
effective Planck scale.
However, and so far, the proposed signatures of black hole production are based on statistical effects 
(rates and angular distributions), not allowing identification on an event by event basis, 
and may lead to large uncertainties \cite{Olinto}.
In this note events with a double bang topology are proposed as an almost background free
signature.
We shall elaborate upon a particular type of event, characterised by the   
production and instantaneous decay of a microscopic black hole (first bang) followed,
at a measurable distance, by the decay of an energetic tau lepton (second bang). 
The characteristics of such events, and the capability of large ground-based cosmic 
ray experiments like Auger~\cite{Auger} or of the
future space based cosmic ray observatories EUSO~\cite{EUSO} and OWL~\cite{OWL} to 
detect them are discussed below.

\section{The first bang: production and decay of microscopic black holes}
\label{sec:bang1}

The interaction of ultra high energy cosmic neutrinos in the atmosphere has been 
pointed out as a good channel for the study of microscopic black hole 
production~\cite{halzen}. 
In the proposed scenario energetic neutrinos ($E_{\nu} \sim 10^6 - 10^{12}$
GeV) interact deeply in the atmosphere (cross-section $\sim 10^3 - 10^7 $ 
pb)
producing microscopic black holes with a mass of the order of the neutrino-parton
center-of-mass energy ($\sqrt{s} \sim 1 - 10 $ TeV). The rest lifetime of these black
holes is so small ($\tau \sim 10^{-27}$~s) that an instantaneous thermal and democratic
decay can be assumed. The average decay multiplicity ($N$) is a function of the parameters 
of the model (Planck mass {$M_D$}, black hole mass $M_{BH}$, number of extra dimension $n$ ) 
and typical values of the order of
$\sim 10$ are obtained in large regions of the parameter space. A large
fraction of the decay products are hadrons ($\sim 75\%$) but there is a non
negligible number of charged leptons ($\sim 10 \%$)~\cite{halzen,Atlas}. 
The energy spectra of such leptons in the black hole centre-of-mass reference 
frame peaks around $M_{BH}/N$. 

The production of tau leptons in black hole decays, in particular the tau
energy spectrum,
was simulated using the CHARYBDIS generator~\cite{Charyb},
while the black holes themselves were produced according to phase-space
and without assuming an explicit $\nu N \rightarrow BH + X$ cross-section, 
as explained in section~\ref{sec:obs}.
Although the CHARYBDIS generator 
was developed to simulate the production and decay of black holes in hadron colliders,
the careful treatment of the black hole decay made it quite useful for the purpose of the
present study. 
Some words about conventions and parameter setting in the CHARYBDIS generator are in order.

The number of extra dimensions $n$ is constrained by existing observational data to be
$n>2$ \cite{Nconstr} (for example $n=1$ is immediately excluded \cite{Hamed} since Newton's 
law holds at solar system scales). We shall take $n=3$ and $n=6$, but the conclusions
apply equally well to higher dimensions. 

At present, in the absence of a quantum theory of gravity, the only way to handle 
theoretically the black hole decay process is to consider black holes masses 
for which quantum gravity effects play 
a negligible role. This is the case for $M_{BH}>M_{min}>>M_D$. We shall however 
relax this constraint and 
consider that a semi-classical description is valid for $M_{min}>M_D$, as usually done in the 
literature. In this study we will consider
$M_{min}=2M_D, 3M_D, 5M_D$ and $10M_D$.
Although the produced black hole could have higher mass than the minimal one 
(depending on the impact parameter and the available centre-of-mass energy of the parton-parton
collision), we work with fixed black hole masses, $M_{BH}=2M_D, 3M_D, 5M_D$ and $10M_D$.
This choice makes our results independent of the uncertainties on the extrapolation of the Parton
Distribution Functions (PDFs) to very high energies and on the computation model for the
black hole production cross-section. Furthermore, it is always possible to convolute the
results, a posteriori, with a given  $M_{BH}$ spectrum.
 
When dealing with black holes, the only conserved parameters (at least in 4 dimensions, $n=0$)
are the total mass, angular momentum and electric charge. There is no lepton
or baryon number conservation~\cite{Teitelboim}, which is sometimes stated by saying that a black hole has 
no hair. The CHARYBDIS 
generator conserves electric charge and allows lepton number violation.
It does conserve baryon number, which however should not be a major drawback \cite{Charyb}. 

We note that there are several different definitions for the Planck mass $M_D$ in 
the literature. Here we shall define it by
\begin{equation}
M_D={(2^{n-3}\pi^{n-1})}^{\frac{1}{n+2}}G_n^{- \frac{1}{2+n}}\,,
\label{mpdef}
\end{equation}
where $G_n$ is Newton's constant on the $n+4$ universe.

CHARYBDIS is interfaced with PYTHIA~\cite{pythia} and HERWIG~\cite{herwig} for standard particle 
hadronisation and decay. The PYTHIA option was selected in this study.
For the termination of the decay, the CHARYBDIS flag KINCUT was set to false, with NBODY=2.

\begin{figure}[hbtp]
\begin{center}
\setlength{\unitlength}{0.0105in}%
\includegraphics[width=0.65\linewidth]{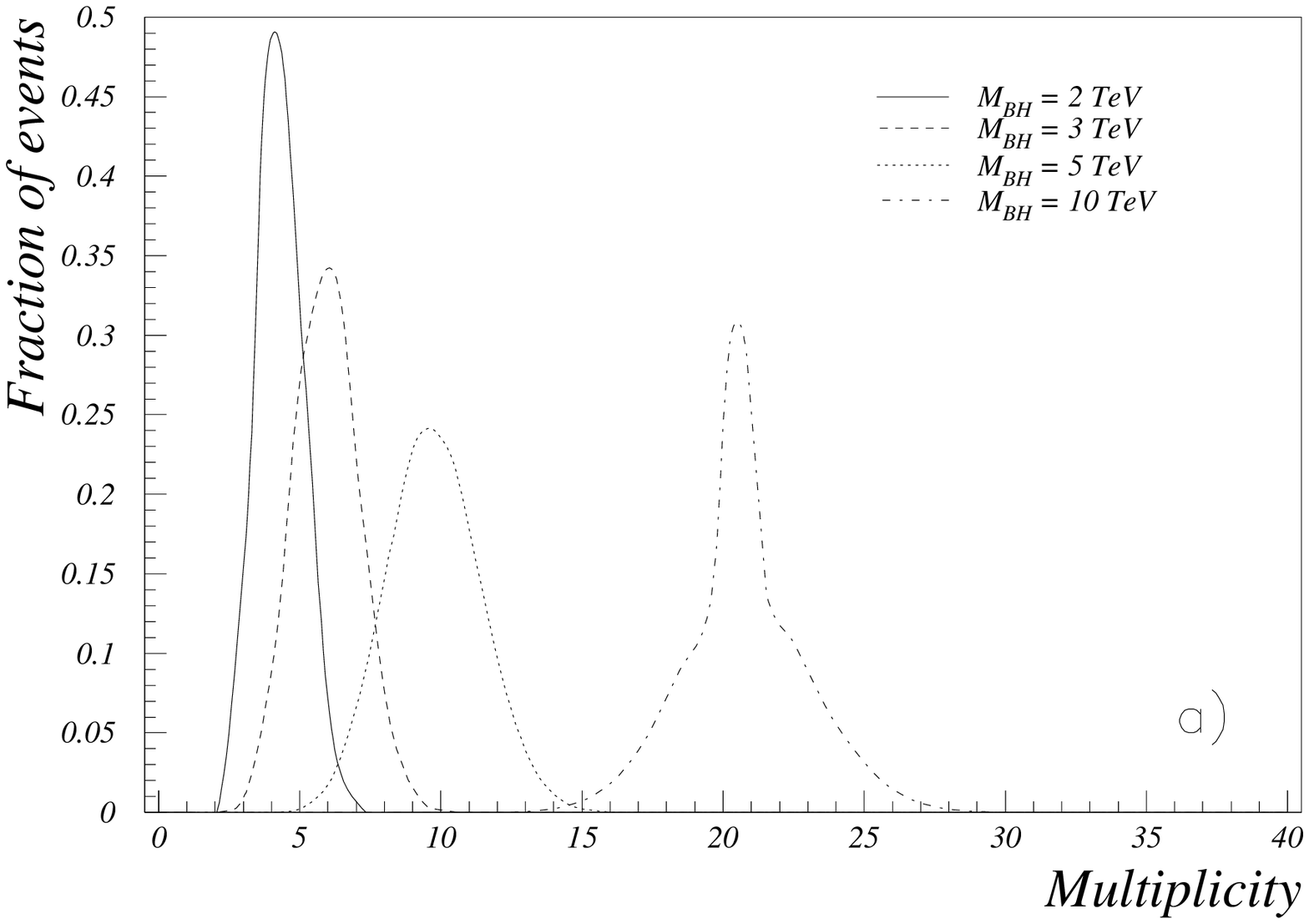}
\includegraphics[width=0.65\linewidth]{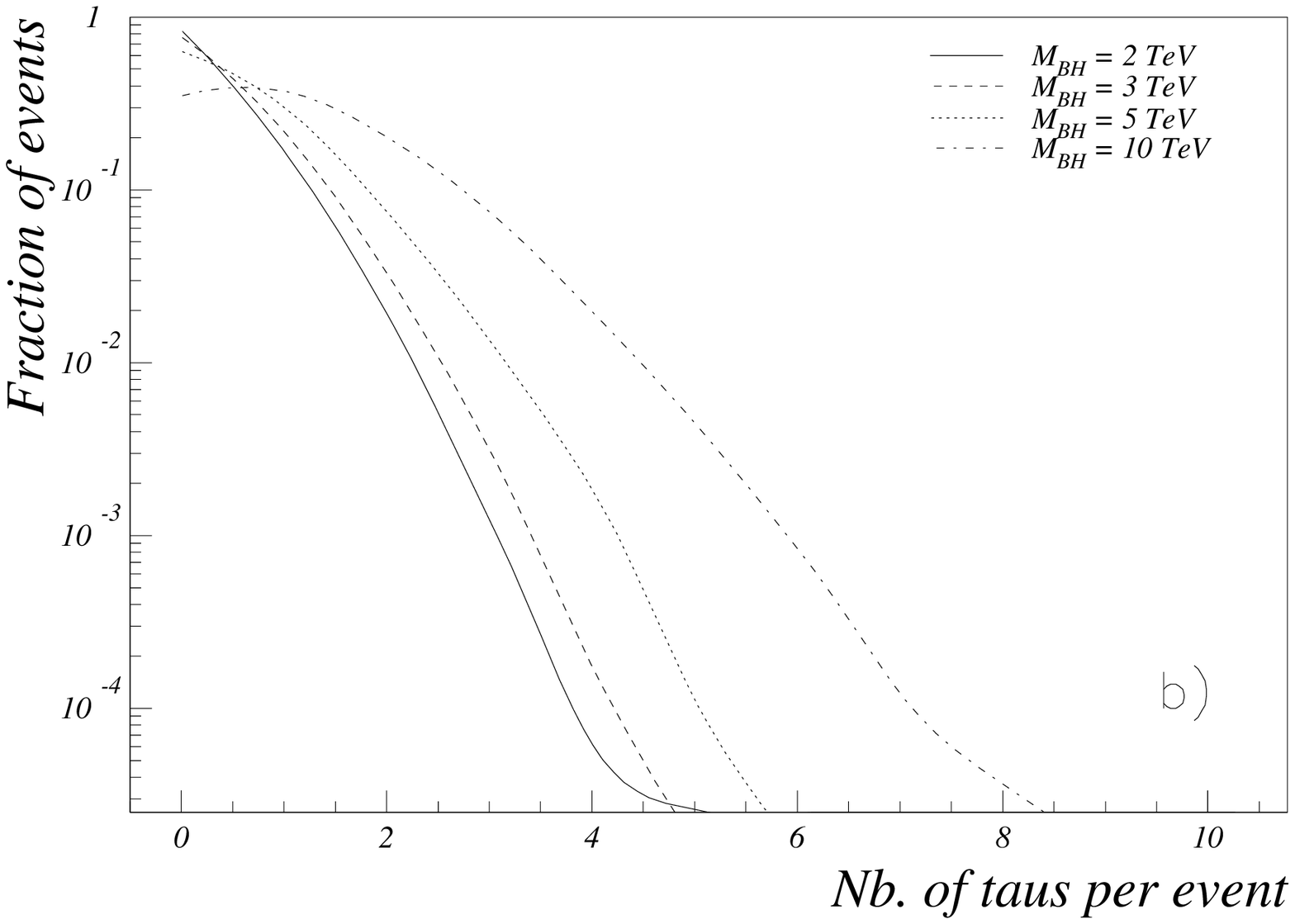}
\includegraphics[width=0.65\linewidth]{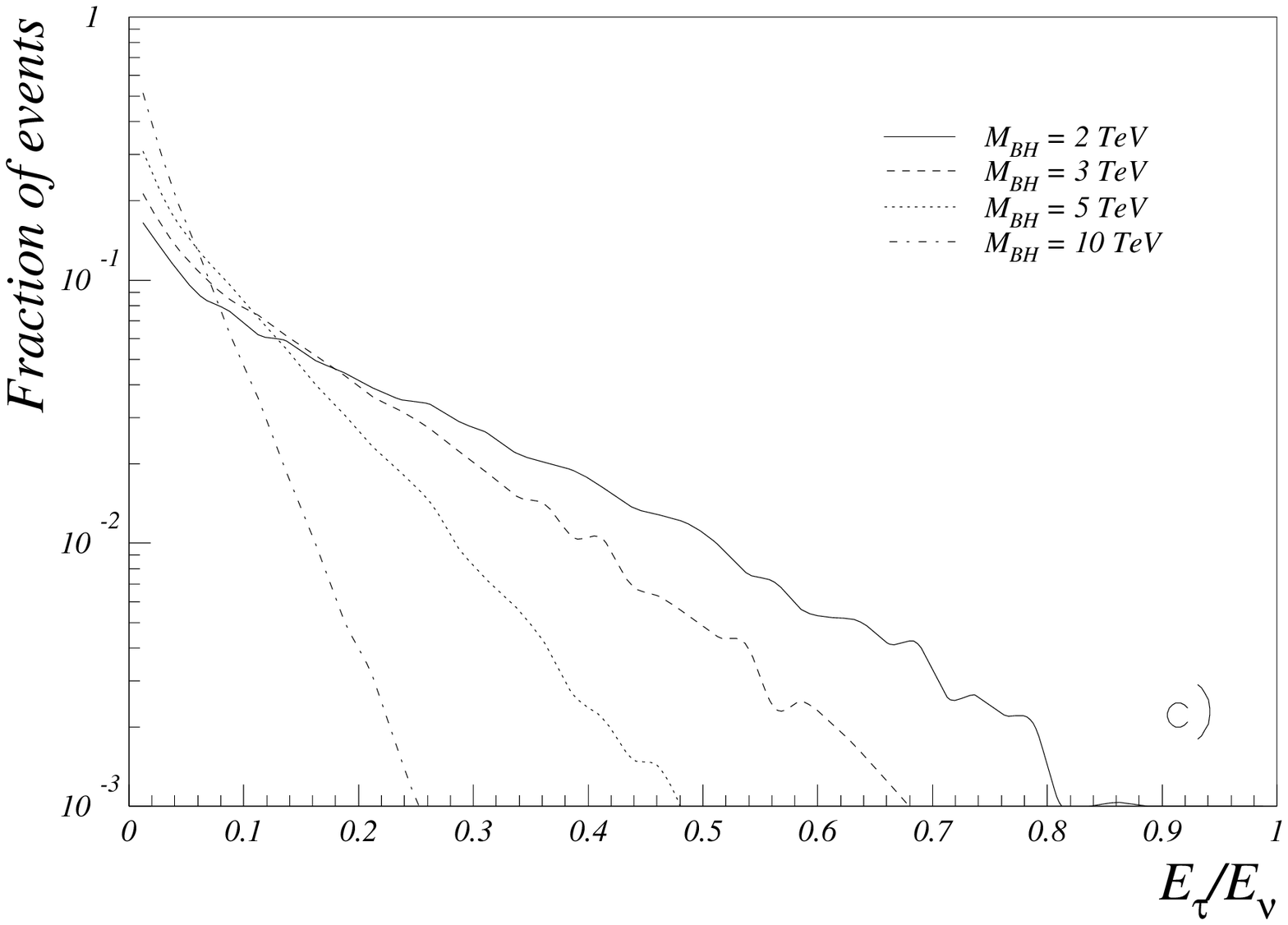}
\end{center}
\caption{
For the scenario ($M_D$=1~TeV, $n$=3) 
and different values of the black hole mass ($M_{BH}$=2,3,5,10 TeV)
are shown: (a) the multiplicity in black hole decays, 
(b) the number of taus per event, (c) the  
fraction of the neutrino energy carried by taus.
}
\label{fig:charyb}
\end{figure}

In figure~\ref{fig:charyb}(a) the multiplicity in black hole decays
is shown for the scenario ($M_D$=1~TeV, $n$=3), for different values of the black hole 
mass, $M_{BH}$. It should be noted that these distributions were, for illustration
purposes, generated at fixed black hole masses, and should in general be convoluted
with the mass spectrum of the produced black holes. While this mass spectrum is
strongly peaked at the lowest possible values, the contribution from the
high mass tail significantly affects the average multiplicitiy. This should
be taken into account when comparing with the literature (e.g. reference~\cite{halzen}).
Figures~\ref{fig:charyb}(b) and (c) show,
for the same ($M_D$, $n$) scenario and
several $M_{BH}$ values, the number of expected 
tau leptons per event and the tau energy spectrum, respectively.  
These distributions have a moderate dependence on the ($M_D$, $n$) parameters and
are mainly determined by the black hole mass, $M_{BH}$.

\section{The second bang: the decay of energetic Tau leptons}
\label{sec:bang2}

Induced tau lepton air-showers have been proposed as a ``golden'' signature
for cosmic neutrino detection, and thus extensively studied~\cite{Fargion,Bottai,dbang-auger}.
In this context, the most interesting property of energetic tau leptons is
their capability of going through ordinary matter, with long decay lengths 
and without important energy losses. In fact, in the relevant energy range, the 
tau interaction length in air is much higher than its decay length, which is
given by $L_{decay} = 4.9$ Km ($E_{\tau}/10^8$ GeV)~\cite{halzen}.
This same property makes taus suitable for tagging microscopic black hole production
in horizontal air shower events.

\begin{figure}[hbtp]
\begin{center}
\includegraphics[width=0.7\linewidth]{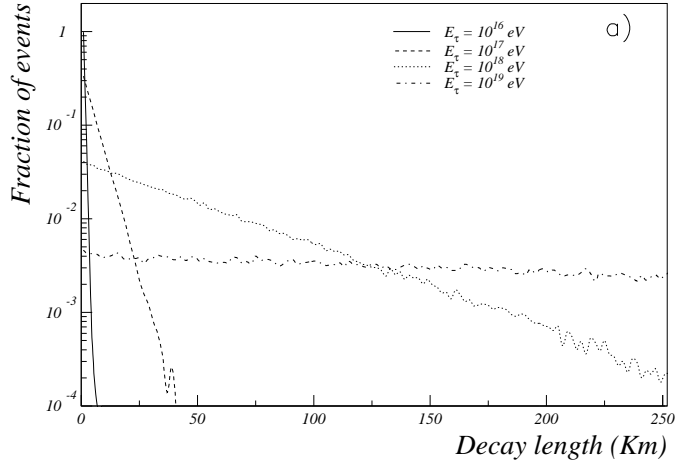}
\includegraphics[width=0.7\linewidth]{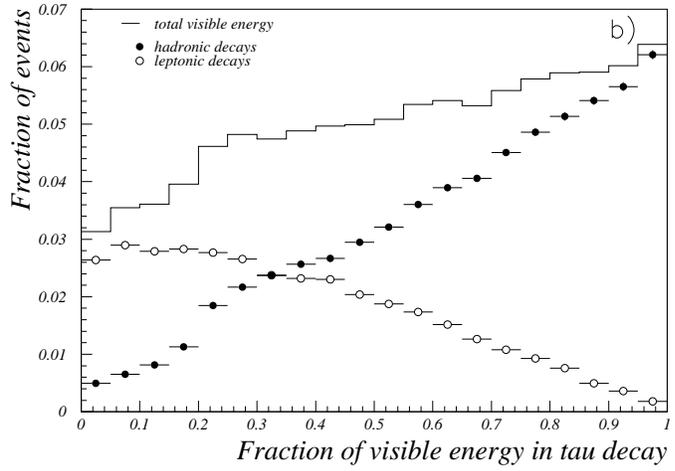}
\end{center}
\caption{Tau decay length for different energy values (a)
and fraction of visible energy in $\tau$ decays (b).}
\label{fig:taudec}
\end{figure}

A detectable second bang can be produced for tau leptons with a decay
length large enough for the two bangs to be well separated, but small enough 
for a reasonable percentage of decays to occur within the field of view.
This is of course determined by the tau energy and, as shown above 
(see figure~\ref{fig:charyb}) tau leptons from black hole decays
are expected to carry of the order of 1/10 of the black hole energy. 
Figure~\ref{fig:taudec}(a) shows, for energies ranging from $10^{16}$ eV to 
$10^{19}$ eV, the distribution of the tau decay lengths. 

Another critical aspect for the detectability of the second bang is 
the visible energy in the tau decay.
Taus can decay leptonicaly ($\sim 34 \%$) or hadronicaly ($\sim 66 \%$).
In both cases, a fraction of the energy escapes detection due to the presence of neutrinos. 
In figure~\ref{fig:taudec}(b) the fraction of energy not associated to neutrinos
in $\tau$  decays is shown. As observed, the average value of this energy, considering all decay
modes, is of the order of 50\%. Hadronic decays produce a large amount of visible
energy, which will be seen as an extensive air shower. For leptonic decays, not only the 
fraction of energy not associated to neutrinos is lower, but also only decays into electrons
originate extensive air showers, leading to observable fluorescence signals.

Summarising, once a first bang is observed, the detectability of the second bang is basically 
determined by the tau lepton energy, which determines both the tau decay length and the 
energy of the second shower. The key characteristics of cosmic ray experiments are thus
the field of view and the energy threshold.

\section{Observation Prospects}
\label{sec:obs}

The observation of double bang events with two showers separated
by tens or hundreds of kilometers 
represents an important challenge to the next generation
of large-scale cosmic rays experiments.
The detection of the fluorescence light produced by the development
of an UHECR shower in the atmosphere is a well suited technique ~\cite{airflu}
to measure the longitudinal shower profile. It has however two severe
limitations: the very low efficiency of the energy conversion to fluorescence
photons, and the atmosphere attenuation length. 
The first constrain restrains the application of such
technique to the detection of very high energy showers in low background
conditions (moonless nights). Threshold energies around  $10^{19}$~eV or
$5~10^{19}$~eV have been quoted for Auger ~\cite{Auger} and EUSO \cite{EUSO},
respectively. The second constrain applies mainly to the Earth-based experiments
(Auger), as the path of the fluorescence photons to space-based experiments
is usually below one attenuation length. 
In space based experiments, not only these attenuation effects are less critical
but also a larger number of black hole induced events is typically expected,
due to the larger acceptance~\cite{OWL-BH}. This makes space-based experiments 
more promising for the detection of the double-bang signature events. 

\begin{figure}[hbtp]
\begin{center}
\includegraphics[width=0.7\linewidth]{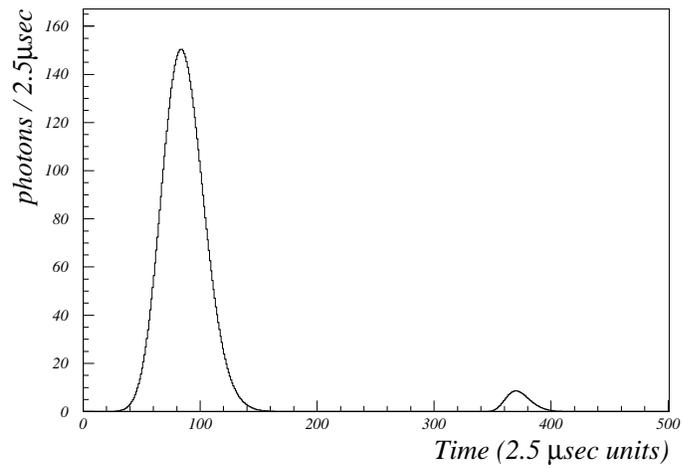}
\end{center}
\caption{
Longitudinal profile of double bang event,
originating from an horizontal incoming neutrino
with an energy of $10^{20}$ eV.
}
\label{fig:longprof}
\end{figure}

In the following paragraphs EUSO will be taken as a case study, without however
making any detailed simulation. 
In fact, double bang events were generated parameterising the shower development
and the atmosphere response, in a model implemented in MATHEMATICA~\cite{math}.
This approach was inspired in the method presented in the description of the
SLAST simulation~\cite{slast}. The GIL parameterisation~\cite{gil} was used 
for the longitudinal  
shower development, while fluorescence yield follows reference~\cite{nagano}.
Attenuation in the atmosphere was included using tables produced with 
LOWTRAN~\cite{lowtran}.
A cross-check with reference plots for EUSO~\cite{redbook} at the 
detector entrance was made and a reasonable agreement (10-20$\%$) 
was obtained.
The results obtained must thus be considered
as an order of magnitude computation. To translate these results to OWL~\cite{OWL} an
increase by a factor of five in expected event rates can be 
considered~\cite{OWL-BH}.   
In figure~\ref{fig:longprof}, the longitudinal fluorescence profile
of a double bang event at the entrance of EUSO is shown as an example.

The observation window for the double bang events is constrained
by geometrical considerations (the two showers must be inside the field of view) 
and by the signal to noise ratio. The reconstruction of the second
shower, which has an energy one order of magnitude lower than the first one,
is a critical issue. However, the energy threshold for this second shower
is only determined by the expected number of signal and background photons
in a very restricted region of the field of view, as the second shower
must be aligned with the direction of the first one.

\begin{figure}[hbtp]
\begin{center}
\includegraphics[width=0.7\linewidth]{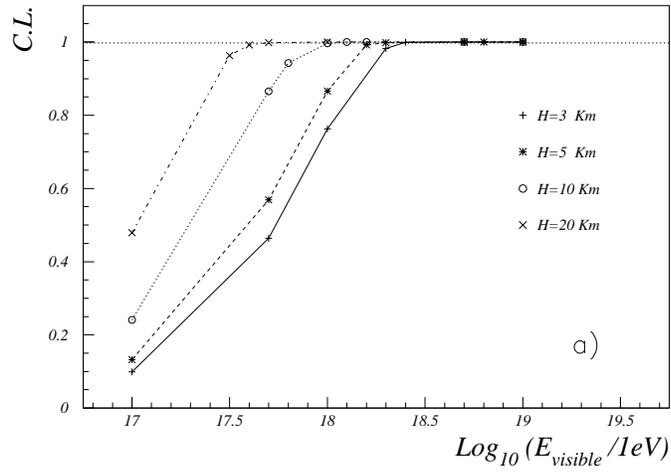}
\includegraphics[width=0.7\linewidth]{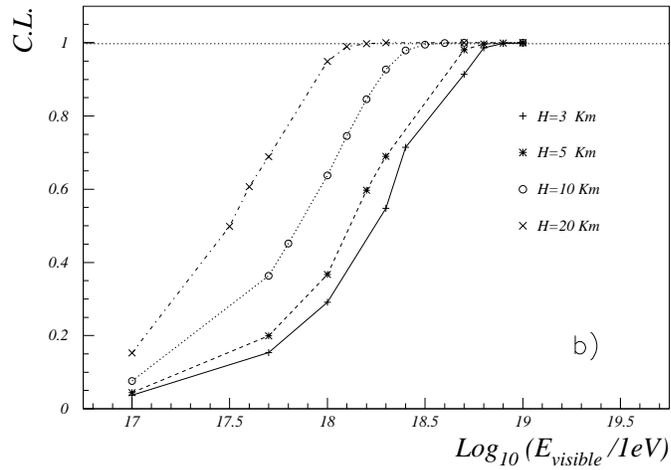}
\end{center}
\caption{
Visibility Confidence Level of the second shower as function of its visible energy,
at a simplified EUSO detector with efficiency (a) 1.0 and (b) 0.1, for
horizontal showers at different heights. The horizontal dotted line shows
the 99.7\% CL ($3\sigma$).}
\label{fig:ConfLevel}
\end{figure}

In figure~\ref{fig:ConfLevel} the Confidence Level (CL) for observing the second
shower is shown as a function of the shower visible energy, for 
horizontal showers at different altitudes. 
The modified frequentist likelihood ratio method~\cite{MFLR},
which takes into account not only the total number of expected signal and background
events but also the shapes of the distributions, was used. 
Signal events were obtained using the method described above.
The number of background photons has been estimated considering an expected background
rate of 300-500 photons/(m$^2$.ns.sr)~\cite{baby} (corresponding to about 0.5-0.7 
photoelectrons per pixel in the EUSO focal surface in a time interval of 
2.5$\mu$s~\cite{redbook}) and assuming a flat distribution.
An ideal photon detection efficiency of 1.0 and a more realistic one of 0.1 were considered.
Horizontal incoming neutrinos at different heights were generated, with the 
horizontal coordinates of the
point of first interaction randomly distributed inside the EUSO field of view.

A CL of 99.7\% (3$\sigma$) was chosen as the criterion of visibility of the second
shower. 
Threshold energies as low as $5 \times 10^{18}$ eV ($1 \times 10^{18}$ eV) 
can be obtained for a photon
detection efficiency of 0.1 (1.0)  and a shower height of 10 Km. The height
is an important parameter, as the number of detected fluorescence
photons increases with the height. 
A factor of 20 variation in the number of detected photons 
can be obtained between two similar showers at low ($~$3 Km) and high ($~$20 Km) height.

To study the visibility of double bang events in EUSO, black holes
produced by horizontal incoming neutrinos
were generated, $\nu N \rightarrow BH + X$, and  
decayed (see section~\ref{sec:bang1}). 
The horizontal coordinates of the point of first interaction 
were randomly distributed within the field of view, 
and the shower height was randomly chosen according to the density profile.
The energy of the shower resulting from the black hole decay (first bang)
was computed and tau leptons originated from this decay were
followed and decayed (second bang).
The visibility of the tau decay shower was established using the CL 
criterion described above.

\begin{figure}[hbtp]
\begin{center}
\includegraphics[width=0.7\linewidth]{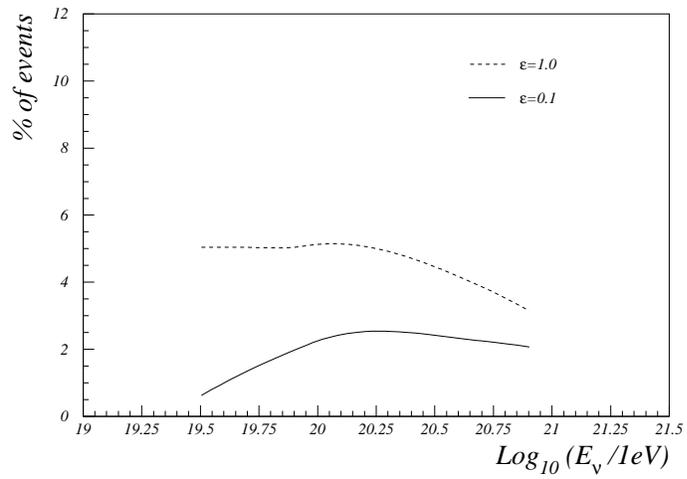}
\end{center}
\caption{
Fraction of events with a first bang within the EUSO field of view that 
also have a visible second bang, as a function of the different primary 
neutrino energy $E_{\nu}$, for ($M_D$=1 TeV,$n=3$,$M_{BH}$=5 TeV), and
for detection efficiencies $\epsilon=$0.1 and $\epsilon=$1.
}
\label{fig:results}
\end{figure}

The fraction of black hole events with an observable second shower 
is shown in figure~\ref{fig:results}, as a function of the primary
neutrino energy, for ($M_D$=1~TeV, $n$=3,$M_{BH}$=5~TeV),
and for detector efficiencies of 1.0 and
0.1. These results take into account
the fraction of events with taus in black hole decays,
the tau energy spectrum and its decay length, the geometrical
acceptance of EUSO and the visibility of the second shower.
For a detector efficiency of 0.1 (1.),
at $E_{\nu} =  10^{20}$~eV, of the order of 2\% (5\%) of the black hole induced events with 
a black hole decay  visible in EUSO (first bang) are expected to have a visible second bang. 
The expected number of double bang events detectable in EUSO is limited at the lower neutrino
energies by the shower visibility and at the highest by the tau lepton decay length.
For relatively low energies the main limiting factor is the energy of the second
bang and, in the figure, the curves corresponding to the two detector efficiencies are far apart.
As the energy increases, the tau decay length increases and the second bang often
escapes detection, and the curve corresponding to an efficiency of 1.0
drops considerably. On the other hand, for an efficiency of 0.1, the favourable effect
of the increase of the second bang energy is still dominant, and the two curves approach.
Different values of $M_{BH}$ can lead to important changes in these results, as the 
black hole decay multiplicity and the tau energy spectrum are strongly dependent on
$M_{BH}$ (see section~\ref{sec:bang1}). The dependence on the ($M_D$,$n$) parameters 
was found to be small.

The rate
of black hole induced events depends strongly on the assumed cosmogenic and extragalactic 
neutrino fluxes. Values ranging
from several tens to hundreds of events per year, for $M_D$ = 1 TeV, have been predicted
for the OWL telescope  \cite{OWL-BH}. In the same reference the acceptance 
of EUSO was estimated to be $1/5$ of OWL's. 

In EUSO, the expected rate of double bang events due to black hole production is therefore
small. However, the observation of just a few such events would be of the utmost
importance, as almost no standard physics process has this signature. The most
dangerous background comes from the $\nu_{\tau}$ regeneration in the Earth
atmosphere ( $\nu_{\tau} N \to \tau X \to  \nu_{\tau} X$)
~\cite{dbang-auger}, which would be,
in itself, a very interesting observation. However, in this type of events
the second shower is expected to be more energetic than the first one.
Other new physics channels may lead to similar
signatures, but the observation of a reasonable number of events 
would give some discriminating power between different models, through the
measurement of the energies of the two showers and of the distance between
them.
 
In the case of Auger the field of view is smaller, and the observable decay
lengths are restricted to about 30 Km 
if the combination of two fluorescence eyes is considered.
On the other hand, the lower energy
threshold  would allow the exploration of a lower
neutrino energy window ($E_{\nu} \sim 10^{19} $ eV) if the expected neutrino fluxes 
and/or the expected 
cross-section are higher than the values usually quoted.  
 
\section{Conclusions}
\label{sec:concl}

The next generation of large cosmic ray experiments (Auger, EUSO, OWL) may have access
to events with a double bang topology, an almost background
free signature. This signature was explored in the framework of the production
of microscopic black holes in the interaction of UHECR in the atmosphere.
EUSO, which has an intermediate field of view between Auger and OWL, was taken
as a case study.

The studies presented in this paper show that it is possible to reconstruct 
second bang showers above an energy threshold which is much lower than the
one required to trigger on the first one. This requires that, once a
very inclined shower trigger occurs, the portion of the field of view 
where the second bang may occur (aligned with the first shower) is also
readout. 
In this conditions, it is shown that the second bang observation
probability, once the first bang is observed, can be of the order of a
few \%.

The predicted rates of double bang events are, in the case of EUSO, 
small and strongly dependent on the
expected neutrino fluxes and on the computed cross-sections,
both with large uncertainties. 
However, the discovery potential of such events is so high
that the next generation of cosmic ray experiments 
should consider them in their
performance studies and in the optimisation of their trigger
and readout systems.



\newpage

\newpage

\newpage

\newpage

\newpage

\end{document}